\documentclass[11pt,a4paper]{article}

\usepackage{graphicx}
\usepackage{epsfig}
\usepackage{appendix}

\usepackage{hyperref}
\usepackage{color}
\usepackage{graphicx,epsf}
\usepackage{bm}
\usepackage{amsmath}

\begin{document}

\begin{center}
{\LARGE{\textbf{Geodesic Acoustic Modes in pair plasmas confined in tokamak magnetic fields}}}\\
\vspace{0.2 cm}
{\normalsize A. Biancalani,} \\
\small{De Vinci Higher Education, De Vinci Research Center, 92 916 Paris La Defense, France
}
\end{center}

\begin{abstract}
This paper is devoted to the derivation of the dispersion relation of the Geodesic Acoustic Mode in pair plasmas, i.e. assuming that ions and electrons have the same mass. Geodesic Acoustic Modes are plasma perturbations playing a crucial role in turbulence regulation, and therefore in the determination of the plasma confinement in tokamaks. Experiments with pair plasmas, like electron-positron plasmas, have been proposed with different kinds of confinements, and aim to study fundamental processes in plasma physics and understanding the formation of the early universe.
\end{abstract}

\section{Introduction}
\label{sec:intro}

Plasmas are ubiquitous in nature. We find them in the stars, in the interplanetary medium, or in the planetary magnetospheres. On earth, we find them whenever there is a ionization process, like in the flames of a fire, or during a lightening of a storm. They are used in industrial processes, and in scientific experiments, like nuclear fusion devices. Most of the examples given here consider plasmas made of ions and electrons, and therefore assume that the mass ratio of the positively and negatively charged particles is several orders of magnitude.
There are plasmas in nature where the mass ratio is unity, like electron-positron plasmas. Electron-positron plasmas are thought to exist around pulsars~\cite{Arons79} and in relativistic jets observed around neutron stars and active galactic nuclei~\cite{Wardle98, Hirotani00, Homan09}.

Experiments with pair plasmas like electron-positron plasmas have been proposed with the goal of studying the fundamental properties of such plasmas, and learn more about the present universe and its formation. Examples are the APEX (a positron–electron experiment) and EPOS (electrons and positrons in an optimized stellarator)~\cite{Stenson19}. In both these examples, the plasma is confined by means of a magnetic field. In the former, a dipole magnetic field is used. In the latter, the magnetic field has both a toroidal and poloidal component. Electron-positron plasmas can exist in tokamak plasmas with runaway electrons in the post-disruption phase~\cite{Helander03}.

The common experiments conducted in magnetic-confinement fusion devices, where plasmas are constituted by mainly ions of hydrogen isotopes and electrons, can help us predicting some of the processes we can find in pair plasmas. For example, we know that density and temperature gradients can yield microinstabilities, which nonlinearly interact forming turbulence. Turbulence is responsible for the enhancement of energy and particle transports from the hot core to the colder edge, thus strongly diminishing the confinement time. Hence, understanding turbulence is crucial if we want to achive a better confinement, which is a mandatory step towards controlled nuclear fusion. A recent theoretical study of turbulence in electron-positron plasmas is shown in Ref.~\cite{Stoneking20}.

Turbulence is often observed in tokamaks in the presence of zonal, i.e. axisymmetric, flows. Zonal flows play a crucial role in regulating turbulence intensity. Two kinds of zonal flows are known in tokamak plasmas: zero-frequency zonal flows~\cite{Hasegawa79, Chen00}, and geodesic acoustic modes (GAMs)~\cite{Winsor68, Sugama06, Sugama08, Zonca08, Qiu09}. Due to their finite frequency, GAMs are affected by Landau damping, whereas zero-frequency zonal flows are mainly damped by collisional damping. GAMs in pair plasmas have been numerically observed in Ref.~\cite{HornStanja19}.

In this paper, we derive the analytical dispersion relation of GAMs in pair plasmas confined in tokamak devices. Like for common plasmas, a kinetic theory is necessary due to the frequency of GAMs, which is of the same order of magnitude of the transit frequency of the particles. The resonances with the passing particles not only determines the intensity of the Landau damping, but also strongly modify the real part of the GAM frequency. Due to the fact that the GAM frequency is much lower than the cyclotron frequency, we can average out the fast gyromotion of the particles around the magnetic field line, thus reducing the 6D kinetic model to a 5D gyrokinetic (GK) model. For our calculation, we invoke the degeneracy of the GAMs with a type of Alfv\'enic instability named beta-induced Alfv\'en eigenmode~\cite{Zonca08}. Due to this degeneracy, we can follow a similar derivation as the one given in Ref.~\cite{Zonca96} for the Alfv\'en eigenmodes. 

We find that in pair plasmas, the GAM frequency is lower but of the same order of magnitude as in common plasmas. The damping rate is also higher than in common plasmas, and the ratio can reach orders of magnitude. We provide the dependence of the frequency and damping rate on the safety factor, and on the plasma temperature. The analytical dispersion relation derived here gives results consistent with the simulations shown in Ref.~\cite{HornStanja19}.

The paper is organized as follows. In Sec.~\ref{sec:model}, the model is described and the regime of validity is described. In Sec.~\ref{sec:GK-resolution}, the GK equation is solved. In Sec.~\ref{sec:disp-rel}, the dispersion relation is given. Finally, in Sec.~\ref{sec:summary}, a summary of the results is provided.

\section{Model equations}\label{sec:model}

The appropriate framework for the study of acoustic modes with the frequency of the order of the transit frequency of the particles, is the GK theory. Here, we refer to the formulation given in Ref.~\cite{Chen91}. The GK theory allows us to include all kinetic effects, yet averaging out the fast gyromotion of the particles spiraling around the magnetic field lines. Thus, the GK model is capable of describing the dynamics of waves with frequency being much lower than the cyclotron frequency. This is the case for acoustic modes, whose frequency is of the order of magnitude of the sound frequency $\omega_s = c_s/R$, with $c_s=\sqrt{T/m}$ being the sound velocity. Thus the ratio of the sound frequency with the cyclotron frequency is $\omega_s/\Omega_i = \epsilon_{asp} \, \rho_s/a$, with $\epsilon_{asp} = a/R$ being the inverse aspect ratio, i.e. the ratio of the tokamak minor and major radius, and $\rho_s$ being the sound Larmor radius. Both $\epsilon_{asp}$ and $\rho_s$ are much smaller than unity for typical tokamak parameters.

The dispersion relation of GAMs in GK theory has been derived, for example,  in Ref.~\cite{Sugama06, Sugama08, Zonca08, Qiu09}, in the case of a large ratio of ion to electron mass. Here, we take Ref.~\cite{Zonca08} as a basis for our description for pair plasmas. Note that, as explained in Ref.~\cite{Zonca08}, the GAM dispersion relation is degenerate with the dispersion relation of low-frequency Alfv\'en modes, which was derived in Ref.~\cite{Zonca96}, if the effect of the density and temperature gradients is neglected. Therefore, we follow the derivation of Ref.~\cite{Zonca96}, just dropping the hypothesis of large mass ratio.

The equilibrium distribution function $F_0$, i.e. the part of the distribution function which is constant in time, is assumed to be Maxwellian (note that from here on, the subscript $s$ refers to the species $s$): 
\begin{equation}\label{eq:Maxwellian}
 F_{0s} = n_0 \Big( \frac{m_s}{2\pi T_s}\Big)^{3/2} \exp\Big( -\frac{m_s (v_\|^2+v_\perp^2)}{2T_s} \Big)
\end{equation}
Here $n_0$ is the local equilibrium density (the same for negative and positive particles due to the local neutrality), and $v_\|$ and $v_\perp$ are respectively the parallel and perpendicular components of the particle velocity with respect to the (equilibrium) magnetic field.
The perturbed particle
distribution function is written as a sum of the adiabatic and non-adiabatic parts:
\begin{equation}\label{eq:adiab-nonadiab}
\delta f_s = - \frac{e_s}{T_s} F_{0s} \widetilde{\delta \phi} \; + \; \delta K_s e^{i L_{ks}}
\end{equation}
where $\delta\phi = \overline{\delta \phi} + \widetilde{\delta \phi}$ is the perturbed scalar potential, written as the sum of the zonal (i.e. flux-surface averaged) and non-zonal (i.e. oscillatory on a flux surface) components. In Eq.~\ref{eq:adiab-nonadiab}, $\delta K_s$ is the non-adiabatic component of the perturbed guiding-center distribution function, and $e^{i L_{ks}}$ is the operator projecting from guiding center coordinates to particle coordinates, with $L_{ks} = (m_s c / e_s B) {\bf (k\times B)\cdot v}$. Note that, in Ref.~\cite{Zonca96}, the nonadiabatic part of the electron perturbed distribution function was assumed to be vanishing, due to the much smaller mass of the electrons with respect to the ions. This is a major difference of this paper with respect to Ref.~\cite{Zonca96}: here we keep the nonadiabatic parts for both positive and negative particles.

We consider here the electrostatic limit of the model equations of Ref.~\cite{Zonca96} (i.e. we neglect the magnetic perturbations), for a homogeneous plasma (i.e. we assume that the equilibrium profiles have no gradients).
The GK equation gives us the evolution of the nonadiabatic component of the distribution function $\delta K$ as a function of the perturbed scalar potential $\delta\phi$:
\begin{equation}\label{eq:GK}
\big( \omega_{tr} \partial_\theta -i (\omega - \omega_d ) \big)_s \delta K_s = -i \frac{F_0 e_s}{T_s} \omega \Big( J_0 \widetilde{\delta \phi} + \Big(\frac{\omega_d}{\omega}\Big)_s J_0 \overline{\delta \phi} \Big) 
\end{equation}
where $\omega_{tr} = v_\|/qR$, $\omega_{ds}(\theta)= \omega_{ds}(0) \, g(\theta)$, with $\omega_{ds}(0)= (k_{\theta_0} m_s c / e B R) (v_\perp^2/2 +v_\|^2)  $, $g(\theta) = \cos\theta + s\theta \sin\theta$, and $(1+s^2\theta^2) = k_\perp^2 / k_{\theta_0}^2$.
Here $k_{\theta_0}$ is the poloidal wave-number, and $\theta$ is the generic ballooning angle.  The Bessel function of the zeroth order, $J_0=J_0(k_\perp \rho_{Li})$, denotes the gyroaverage operator.

The quasi-neutrality (QN), is derived by equating the densities of the ions and the
electrons: $\sum_s e_s \; \delta n_s = 0$. By splitting the adiabatic and non-adiabatic components, this equation is written as:
\begin{equation}\label{eq:qn-0}
\Big( 1+\frac{1}{\tau}\Big) \widetilde{\delta \phi} = \frac{T_i}{n_0 e} \Big( \langle J_0 \delta K_i \rangle -  \langle J_0 \delta K_e \rangle \Big)
\end{equation}
where $\tau = T_e/T_i$ and the angular brackets denote the integration in velocity space. Alternatively, we can cast it in the form of a vorticity equation. For its derivation, we start with the QN equation written in the form of a GK-Poisson law (see Ref.~\cite{Bottino15} and references therein):
\begin{equation}
\sum_s  \left\langle e_s J_0 \; \delta f_s + \nabla \cdot \frac{m_s c^2}{B^2} f_{0s} \nabla_\perp \delta \phi\right\rangle = 0
\end{equation}
Taking the time derivative and putting this equation in a continuity form, we obtain the vorticity equation:
\begin{equation}
\sum_s (n_s m_s) \frac{c^2}{B^2} \frac{\partial}{\partial t} \nabla_\perp^2 \delta \phi + \nabla \cdot \delta {\bf J}_G = 0
\end{equation}
with:
\begin{equation}
\delta {\bf J}_G = \sum_s \left\langle e_s \delta f_s \dot{\bf R}_0 \right\rangle 
\end{equation}
with $\dot{\bf R}_0$ being the unperturbed velocity. Now taking again the time derivative and considering the flux-surface-averaged component in the electrostatic limit, we obtain~\cite{Zonca96}:
\begin{equation}\label{eq:vorticity}
\frac{\omega^2}{v_A^2}  k_\perp^2 \overline{\delta\phi} = \overline{
\left\langle \sum_s \frac{4\pi e_s}{c^2} J_0 \omega \omega_{ds} \delta K_s \right\rangle }
\end{equation}
with $v_A = B_0/\sqrt{4\pi n_0 \sum_s m_s}$ being the Alfv\'en velocity (here the masses of both positive and negative particles give non-negligible contribution to the denominator of the Alfv\'en velocity, differently from ordinary plasmas, where electrons are negligible).
Note that, so far, no hypothesis has been made regarding the asymptotic expansion of the fields and equations. Therefore, our two main model equations, Eqs.~\ref{eq:GK} and \ref{eq:vorticity} are valid for a variety of spatial and temporal scales.

Now, before proceeding further, we introduce the ordering of the GAM dynamics, which will be applied to these equations. 
One of the first properties of GAMs we know from experiments and numerical simulations, is that they have nearly $m=0$ electric field, which means $\widetilde{\delta \phi} \ll \overline{\delta \phi}$.
Moreover, as GAMs are sound oscillations, the expansion of the GK equation is done with the smallness parameter $\epsilon$ which measures the ratio $\omega_{di} / \omega \propto k_r\rho_i$, i.e. the square root of the plasma temperature (note that in the original reference, Ref.~\cite{Zonca96}, which treats electromagnetic perturbations, we have $\epsilon = \beta^{1/2}$). To the zeroth order, i.e. for temperature going to zero, we predict no GAM oscillation of the distribution function, i.e. $\delta K_s^{(0)}=0$. Therefore, the dynamics of the GAM is given by the resolution for the higher order component, $\delta K_s^{(1)}$. 

For the expansion of the scalar potential, we quantify its maximal ordering as $\widetilde{\delta \phi}/\overline{\delta \phi} \sim \mathcal{O}(\epsilon)$.
Therefore, the decomposition of $\delta \phi$ is described by $\delta \phi = \delta \phi^{(0)} + \delta \phi^{(1)} +$ ..., with $\delta \phi^{(0)} = \overline{\delta \phi} \sim \mathcal{O}(1)$ and $\delta \phi^{(1)} = \widetilde{\delta \phi}^{(1)}\sim \mathcal{O}(\epsilon)$, etc. We take $ \overline{\delta \phi}$ as reference, which is ordered for definition as $\mathcal{O}(1)$.
 
We also recognize that the the ballooning angle $\theta$ can be decomposed in a short-scale angle $\theta_0 \sim \mathcal{O}(1)$, which is the poloidal angle of the tokamak geometry, and a long-scale angle $\theta_1 \sim \mathcal{O}(\epsilon^{-1})$ (this means that we are considering radially localized modes).
As a consequence, the derivatives become $\partial_{\theta_0}\sim \mathcal{O}(1)$ and  $\partial_{\theta_1}\sim \mathcal{O}(\epsilon)$. 
We can now write $g(\theta)$ explicitly in term of the poloidal angle $\theta_0$:  $g(\theta_0) = \cos\theta_0 + s\theta_1 \sin\theta_0$, and $s\theta_1 = k_\perp / k_{\theta_0}$~\cite{Cheng85,Pegoraro86}. This separation of scales, is crucial especially for MHD modes like Alfv\'en instabilities or tearing modes.
For GAMs, we take the hypothesis of $k_\perp / k_{\theta_0} \gg 1$ and therefore we substitute everywhere $s\theta_1 \rightarrow k_\perp / k_{\theta_0} $. This hypothesis of mode radial localization, $k_\perp / k_{\theta_0} \gg 1$, helps us rewriting $\omega_{d,s}(\theta)\simeq \hat\omega_{d,s}\sin\theta_0$, with $\hat\omega_{d,s}= (k_\perp m_s c / e_s B R) (v_\perp^2/2 +v_\|^2)  $. Having separated the scales, we can now focus on the rest of the paper on the $\theta_0$ behavior. For simplicity, from now on, we will drop the subscript ${}_0$, and $\theta$ will refer to the poloidal angle $\theta_0$.

\section{Expansion of the GK equations}
\label{sec:GK-resolution}

For the expansion of the GK equation of each species, we follow the same steps of Ref.~\cite{Zonca96}. The difference is that, here, we treat both species in the same way, whereas in Ref.~\cite{Zonca96} the negative particles (the electrons) were treated adiabatically.
We start the expansion of the GK equation for each species in the smallness parameter $\epsilon \sim \omega_d / \omega$.
As $ \overline{\delta \phi}$ is defined as $\mathcal{O}(1)$, then the terms on the RHS of the GK equation are ordered both as $\mathcal{O}(\epsilon)$. We consider now the QN equation, which sets the order of the lowest non-zero order of $\delta K$, i.e. $\delta K^{(1)}$, to the scalar potential (as we discussed $\delta K^{(0)}$ being null, in the previous section). It follows that the second term on the LHT, considered with its zeroth component for the sake of the ordering of the equation, has maximal ordering $\mathcal{O}(1)$. 
The drift frequency $\omega_d$ (which does not depend on the species mass), on the other hand, is smaller than the GAM frequency with the ordering $\omega_d / \omega \sim \mathcal{O}(\epsilon)$.
Regarding the first term on the LHS, we know from the MHD theory that the GAM frequency is of the order of the sound frequency, which is of the order of the ion transit frequency.
In summary we have $\omega_d / \omega \sim \mathcal{O}(\epsilon)$, and $\omega / \omega_{tr} \sim \mathcal{O}(1)$. Therefore, the GK equation is ordered as:
\begin{equation}\label{eq:GK-ions}
\big( \underbrace{\omega_{tr,s} \partial_\theta}_\text{$\sim \mathcal{O}(1)$} -i (\underbrace{\omega}_\text{$\sim \mathcal{O}(1)$} - \underbrace{\omega_{d,s}}_\text{$\sim \mathcal{O}(\epsilon)$} ) \big) \delta K_s = - i \frac{F_0 e}{T_s} \omega \Big(\underbrace{\widetilde{\delta \phi}}_\text{$\sim \mathcal{O}(\epsilon)$} + \underbrace{\Big(\frac{\omega_{d,s}}{\omega}\Big) \overline{\delta \phi}}_\text{$\sim \mathcal{O}(\epsilon)$} \Big) 
\end{equation}
We start the resolution by taking the flux-surface average of Eq.~\ref{eq:GK-ions}. Due to the fact that all terms, but the second on the left-hand-side (LHS), are oscillatory, this yields:
\begin{equation}\label{eq:GK-ions-0}
 \omega\overline{\delta K_s}= 0 
\end{equation}
to all orders. We also recall that $\delta K_s^{(0)}=0$ for both components (zonal and oscillatory), as discussed above.

Passing now to the next order, i.e. $\mathcal{O}(\epsilon)$, we have 
\begin{equation}\label{eq:GK-ions-1}
\big( \omega_{tr,i} \partial_\theta -i \omega  \big) \delta K_s^{(1)} = -i \frac{F_0 e}{T_s} \omega \Big( J_0 \widetilde{\delta \phi}^{(1)} + \Big(\frac{\omega_{d,s}}{\omega}\Big) J_0 \overline{\delta \phi}^{(0)} \Big) 
\end{equation}
We now proceed to the resolution of Eq.~\ref{eq:GK-ions-1} for $\delta K_s^{(1)}$. For the resolution, we first need to calculate the inversion of the operator of the LHS. To this aim, we write Eq.~\ref{eq:GK-ions-1} in a matrix form. We start by writing $\overline{\delta K_s}^{(1)} = \delta K_0$ $(=0)$ and $\widetilde{\delta K_s}^{(1)} = \delta K_{s,\cos} \cos\theta_0 + \delta K_{s,\sin} \sin\theta_0$, and a similar notation is used for $\delta\phi$. For simplicity, we drop the index $s$ denoting the species from here on, in this section. We obtain:

\begin{equation}
\left( \begin{array}{c c c}
-i\omega     &    0      & \omega_{tr} \\
0            & -i\omega  & 0\\
-\omega_{tr} &    0      & -i\omega
\end{array}
\right)
\left( \begin{array}{c} \delta K_{\cos}\\   \delta K_0  \\  \delta K_{\sin}  \end{array}
\right) 
= 
\hat{F}
\left( \begin{array}{c c c}
1     &    0  & 0 \\
0     &        0            & 0 \\
0     &    \hat\omega_d/\omega  & 1
\end{array}
\right)
\left( \begin{array}{c} \delta \phi_{\cos}\\   \delta \phi_0  \\  \delta \phi_{\sin}  \end{array}
\right) 
\end{equation}
with $\hat{F}=-iF_0 e \omega / T$. Now, by calling $M$ the matrix on the LHS, we can calculate the inverse of $M$ as:
\begin{equation}
M^{-1} =  \frac{-i}{\omega (\omega^2- \omega_{tr}^2)}\left( \begin{array}{c c c}
-\omega^2     &    0      & i \omega \omega_{tr} \\
0             & - \omega^2 + \omega_{tr}^2  & 0\\
-i \omega \omega_{tr} &    0      & -\omega^2
\end{array}
\right)
\end{equation}
Therefore we get:
\begin{equation}
\left( \begin{array}{c} \delta K_{\cos}\\   \delta K_0  \\  \delta K_{\sin}  \end{array}
\right) 
= 
\frac{-i\hat{F}}{\omega (\omega^2- \omega_{tr}^2)}
\left( \begin{array}{c c c}
-\omega^2     &    + i \omega_{tr}\hat\omega_d &i \omega \omega_{tr} \\
0     &        0            & 0 \\
-i \omega \omega_{tr}     &    - i\omega \hat\omega_d  & -\omega^2
\end{array}
\right)
\left( \begin{array}{c} \delta \phi_{\cos}\\   \delta \phi_0  \\  \delta \phi_{\sin}  \end{array}
\right) 
\end{equation}
This can be written as \cite{Zonca96}:
\begin{eqnarray}\label{eq:GK-ions-final}
\left\{
\begin{aligned}
\delta K_{\cos} & = & \frac{-i F_0 e \omega }{T (\omega^2- \omega_{tr}^2)}
\left( i \omega  \delta \phi_{\cos} + \underline{\omega_{tr} \delta \phi_{\sin}} + \underline{\omega_{tr} 
\frac{\hat\omega_d}{\omega} \delta \phi_0} \right) \\
\delta K_{\sin} & = &  \frac{-i F_0 e \omega }{T (\omega^2- \omega_{tr}^2)}
\left( - \underline{\omega_{tr}  \delta \phi_{\cos}} + i \omega \delta \phi_{\sin} + 
i \hat\omega_d \delta \phi_0 \right)
\end{aligned}
\right.
\end{eqnarray}
This is the desired solution of the GK equation for each species. The next step, is to put $\delta K_s$ into the vorticity equation. Before proceeding further, note that here, the underlined terms, are those which will be killed by the integration in velocity space, because odd (this assumes that $F_0$ is even, which is the case for the hypothesis of being a Maxwellian, see Eq.~\ref{eq:Maxwellian}).

\section{Resolution of the vorticity equation}
\label{sec:disp-rel}

The last step of the derivation of the dispersion relation of GAMs is the resolution of the vorticity equation, Eq.~\ref{eq:vorticity}.


\begin{figure}[b!]
\begin{center}
\includegraphics[width=0.46\textwidth]{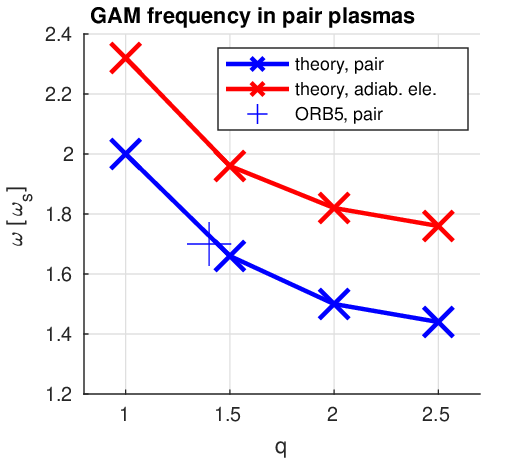}
\includegraphics[width=0.48\textwidth]{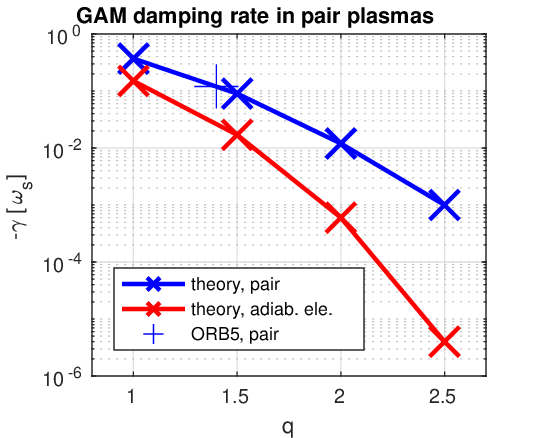}
\caption{\label{fig:omegagamma_q}
GAM frequency (left) and damping rate (right). The theory of pair plasmas is compared with the theory of a plasmas with adiabatic electrons. Moreover, the result of a simulation of a pair plasma given in Ref.~\cite{HornStanja19} is also shown.
}
\end{center} 
\end{figure}

\begin{equation}
 \frac{\omega^2}{v_A^2}  k_\perp^2 \delta\phi^{(0)} = \overline{ 
\left\langle \sum_s \frac{4\pi e}{c^2} \omega \omega_{d} \delta K_s \right\rangle}= 
\left\langle \sum_s \frac{2\pi e}{c^2} \omega \hat\omega_{d,s} \; \delta K_{s,\sin} \right\rangle
\end{equation}
When considering the several terms of $\delta K_s$ in Eq.~\ref{eq:GK-ions-final}, we note again the the underlined terms are killed by integration in velocity space, because odd (again, we assume here $F_0$ being even). 
Of the two remaining terms of $\delta K_{\sin}$ in Eq.~\ref{eq:GK-ions-final}, the first is proportional to the electric charge $e_s$, whereas the second does not depend on the electric charge. Therefore, if we assume that the two spcies have the same temperature and mass, when summing up $\delta K_{\sin}$ for both species with opposite charge, the term in $\delta \phi_{\sin}$ disappears, whereas the term in $\delta \phi_0$ becomes double.
We obtain the following equation:
\begin{equation}
\frac{\omega^2}{\omega_t^2} + \frac{q^2}{4 n_0}  \chi = 0
\end{equation}
where
\begin{equation}
\chi = \frac{8 n_0}{\pi^{1/2}}  \int_0^\infty d\zeta_\perp \int_{-\infty}^\infty d\zeta_\|  \; \zeta_\perp \frac{\exp(-\zeta_\|^2-\zeta_\perp^2)}{\zeta_\|^2-\zeta_{\|res}^2} \zeta_{\|res}^2 \Big(\frac{\zeta_\perp^2}{2} + \zeta_\|^2 \Big)^2 = 4 n_0 \zeta_{\|res} F(\zeta_{\|res})
\end{equation}
where the integration in velocity space has been performed in cilyndrical coordinates $\int d^3 v = 2\pi \int_0^\infty d\zeta_\perp \, \zeta_\perp \int_{-\infty}^\infty d\zeta_\|$, with $\zeta= v/v_t$, $v_t = \sqrt{2T/m}$, $ \zeta_{\|res}= v_{\|res}/v_t$, $v_{\|res} = qR \omega$. 
The complex frequency is normalized as $z=\omega/\omega_t$, and  $F$ is defined by:
\begin{equation}
F(z) = z\Big(z^2+\frac{3}{2}\Big) + \Big(z^4 + z^2 + \frac{1}{2}\Big) \mathcal{Z}(z) 
\end{equation}
We have also introduced the plasma dispersion function $\mathcal{Z}(z) = \pi^{-1/2}\int_{-\infty}^\infty e^{-y^2}/ (y-z) dy$.

Finally, the result is the desired dispersion relation of GAMs, in implicit form in the complex plane:
\begin{equation}\label{eq:GAM-disp-rel}
\boxed{ z + q^2  F(z) = 0 }
\end{equation}

In Fig.~\ref{fig:omegagamma_q}, the solution of Eq.~\ref{eq:GAM-disp-rel} is shown for different values of $q$, and compared with the analytical theory of GAMs in ordinary plasmas as taken from Ref.~\cite{Zonca96,Zonca08}. The numerical result obtained with ORB5 and published in Ref.~\cite{HornStanja19} is also shown, and found to be in good agreement with the theory.

\section{Summary and conclusions}
\label{sec:summary}

The dynamics of pair plasmas is an example of a problem of fundamental plasma physics which can help sheding light on the nature of plasmas. A typical example of pair plasma is the plasma of electrons and positrons. In this case, the positive particles have the same mass of the electrons but opposite charge. Waves and instabilities can be studied in electron-positron plasmas, with differences and analogies with ordinary plasmas of ions and electrons. The theoretical models can be of interest for understanding the dynamics of astrophysical environments like around pulsars, neutron stars and active galactic nuclei. Moreover, theoretical models can be used to model experiments, like 
APEX and EPOS.

In this paper, we investigate the theory of electron-positron plasmas by considering the example of magnetized plasmas in a periodic equilibrium like a tokamak. Those charged particles performing closed trajectories around the machine, will interact with waves in a periodic way. If their transit frequency is of the order of the wave frequency, resonances can strongly modify the wave dispersion relation, thus requiring a kinetic theory for a proper description. Due to the low frequency of interest here, with respect to the cyclotron frequency, we can gyro-average out the fast gyromotion of the particles around the magnetic field lines, decreasing the dimensionality of our kinetic model from 6D in phase space to 5D. This is known as gyro-kinetic theory.

Geodesic acoustic modes (GAM) were observed in numerical simulations of electron-positron plasmas with the gyrokinetic code ORB5~\cite{HornStanja19}. Here, we have derived a dispersion relation of GAMs following the scheme of the calculation given in Ref.~\cite{Zonca96,Zonca08} (i.e neglecting the finite Larmor radius and finite orbit width of the particles). We confirm the existence of GAMs in electron-positron plasmas confined in a tokamak magnetic field, and we investigate the dependence on the safety factor, evaluating the ratio of the poloidal and toroidal components of the magnetic field. Both frequency and Landau damping rate are calculated, and found to decrease with the increasing safety factor. The comparison with ordinary plasmas of ions and electrons shows that the frequency has lower values, but of the same order of magnitude, whereas the damping rate has higher values, which can be different of orders of magnitude for increasing values of safety factors. We also compare the analytical results with the published numerical simulations, finding a good agreement.

This paper serves as a proof-of-principle of how the gyrokinetic theory can be used to study the dynamics of waves and instabilities in non-ordinary plasmas. Possible extensions include the study of the dependence of the GAM dynamics on the temperature and mass ratio of the two species, inclusion of impurities, inclusion of an energetic particle species, or the study of waves and instabilities of other nature.

\section*{Acknoledgements}

This work has been carried out within the framework of the EUROfusion Consortium, funded by the European Union via the Euratom Research and Training Programme (Grant Agreement No 101052200 — EUROfusion). Views and opinions expressed are however those of the author(s) only and do not necessarily reflect those of the European Union or the European Commission.  Neither the European Union nor the European Commission can be held responsible for them. Discussions with Zhiyong Qiu during the visit at ASIPP-Hefei in Spring 2026 are gratefully acknowledged. This paper was written in Trizay (Charente-Maritime). Thanks to Steven Wilson for the vibe.

%
%
%
%
%
%
%

\end{document}